\documentclass[aps,prl,twocolumn,superscriptaddress,showpacs,amsmath,amssymb,floatfix]{revtex4-2}

\usepackage{graphicx} 
\usepackage{dcolumn}  
\usepackage{bm}       
\usepackage{braket}   
\usepackage{xcolor}   
\usepackage{hyperref} 

\begin{document}

\title{Non-Equilibrium Model Selection via Finite-Time Thermodynamics}

\author{Masayuki Ohzeki}
\email{mohzeki@tohoku.ac.jp}
\affiliation{Graduate School of Information Sciences, Tohoku University, Sendai 980-8579, Japan}
\affiliation{Department of Physics, Institute of Science Tokyo, Tokyo 162-8601, Japan}
\affiliation{Research and Education Institute for Semiconductors and Informatics, Kumamoto University, Kumamoto 860-8555, Japan}
\affiliation{Sigma-i Co., Ltd., Tokyo 108-0075, Japan}

\date{\today}

\begin{abstract}
Information criteria such as WAIC and WBIC extend model selection to singular learning machines, but they are usually derived for equilibrium posteriors.
We formulate a finite-time analogue of WBIC by replacing the equilibrium posterior with an effective ensemble generated by learning dynamics under a resource constraint.
When this ensemble admits an analytic effective potential, singular learning theory yields a resource-dependent real log canonical threshold.
The resulting estimator gives a computable thermodynamic contribution to time-bounded MDL and identifies the finite-time singular complexity relevant to the structural information measured by epiplexity.
\end{abstract}

\maketitle

\textit{Introduction.---}
The principle of parsimony, or Occam's razor, is central to statistical modeling and machine learning. 
Information criteria such as the Akaike Information Criterion (AIC) \cite{Akaike1974} and the Bayesian Information Criterion (BIC) \cite{Schwarz1978} formalize this principle by balancing the goodness of fit against model complexity. 
In a statistical-physical language, the goodness of fit is an energy-like quantity, while the model complexity is associated with the volume, or entropy, of the parameter space compatible with the data.
These metrics are often given as static properties of the ``optimal'' model found at the global minimum of a loss landscape, or of the equilibrium posterior around that minimum.
In this sense, classical information criteria are equilibrium quantities from the perspective of finite-time statistical physics.

However, the validity of the simplest forms of AIC and BIC is limited to regular statistical models, where the Fisher information matrix is positive definite. 
They fail in singular models, such as mixture models, hierarchical Bayesian models, and deep neural networks, where parameters are non-identifiable and the Fisher information matrix is rank deficient. 
To overcome this limitation, Watanabe established singular learning theory based on algebraic geometry \cite{Watanabe2009}. 
This framework introduced generalized metrics: the Widely Applicable Information Criterion (WAIC) \cite{Watanabe2010} and the Widely Applicable Bayesian Information Criterion (WBIC) \cite{Watanabe2013}. 
These criteria characterize complexity using the real log canonical threshold (RLCT), denoted by $\lambda$, which is derived from the local algebraic structure of the loss function via Hironaka's resolution of singularities \cite{Hironaka1964,Hironaka1964b}. 

Despite this mathematical sophistication, the standard derivations still retain an equilibrium assumption. 
They assume that the learning agent can access the target posterior distribution through sufficiently long sampling, or that the relevant optimum has been reached by sufficiently long optimization. 
This perspective overlooks a critical constraint: the finite nature of computational resources, denoted here by $R$. 
In modern applications, such as training deep neural networks or solving combinatorial optimization problems via simulated or quantum annealing, the system rarely reaches the true ground state or the exact posterior within the allocated time and resources. 
The observed model is instead the product of a non-equilibrium process, possibly trapped in local minima, metastable regions, or transient states.
Here the finite resource $R$ denotes not only an abstract amount of computation but also the protocol used to generate the learned state.  Concretely, in the formulation below, $R$ specifies a transition kernel $K$ (or an algorithm such as stochastic-gradient Langevin dynamics), its hyperparameters, and a finite time horizon $T$.  Thus finite-resource quantities are properties of the endpoint ensemble characterized by the pair $(K,T)$.

Recently, the concept of ``epiplexity'' has been introduced to address the disconnect between static information measures and dynamic, computationally bounded learning processes \cite{Finzi2026}. 
Unlike traditional Kolmogorov complexity, epiplexity focuses on the structural information extracted by a bounded observer.
In the minimum-description-length formulation of Ref.~\cite{Finzi2026}, a time-bounded predictor is chosen by minimizing a two-part description length, while epiplexity corresponds to the model, or structural, part of the selected code and time-bounded entropy corresponds to the residual predictive loss.
In schematic form, the relevant objective is a quantity such as $-\log P(x|M)+|M|$ subject to $\mathrm{Cost}(x,M)\le R$, where $M$ denotes a model or hypothesis class.
To connect this viewpoint with statistical learning theory, we focus on the model-conditioned stochastic complexity that refines the data-given-model term in the MDL objective.
For regular models this term reduces to the familiar BIC penalty, whereas for singular models it is governed not by the parameter count but by the RLCT.
Watanabe's WBIC is important here because it turns this singular free energy into a tempered expectation, avoiding the explicit resolution of singularities.
We extend this computable free-energy estimator from an equilibrium posterior to a finite-time effective ensemble generated by the learning dynamics.
When the endpoint distribution of a smooth noisy learning process can be represented by an analytic effective potential, the same resolution-of-singularities logic yields a resource-dependent RLCT and hence a finite-time free energy.
This gives a precise sense in which the structural part of a time-bounded description length can contain a time-dependent effective complexity.

\textit{Regular models.---}
Let $D_N=\{x_i\}_{i=1}^N$ be the observed data and let $\theta$ be a $k$-dimensional parameter. 
We define the empirical loss per datum and the corresponding extensive energy as
\begin{equation}
    L_N(\theta)=-\frac{1}{N}\sum_{i=1}^{N}\ln p(x_i|\theta),
\end{equation}
and $E_N(\theta)=N L_N(\theta)$.
This scale distinction is useful: AIC corrects the per-sample predictive loss, whereas BIC estimates an extensive Bayesian free energy.
The maximum likelihood estimator $\hat{\theta}$ is the empirical ground state. 
For a regular model, its fluctuation around the population optimum satisfies
$    \sqrt{N}(\hat{\theta}-\theta^*) \Rightarrow \mathcal{N}(0,H^{-1})$,
where $H$ is the Fisher information matrix.
Write $L_{\mathrm{train}}(\hat{\theta})=L_N(\hat{\theta})$ for the empirical loss and
$L_{\mathrm{gen}}(\hat{\theta})=\mathbb{E}_{x\sim q}[-\ln p(x|\hat{\theta})]$ for the population loss on an independent sample.
We then find
\begin{equation}
    \mathbb{E}_{D_N}\!\left[
    L_{\mathrm{gen}}(\hat{\theta})
    -L_{\mathrm{train}}(\hat{\theta})
    \right]
    \approx \frac{k}{N}.
\end{equation}
AIC cancels this bias by adding $2k$ to the deviance $2N L_{\mathrm{train}}(\hat{\theta})$; thermodynamically, this is the equipartition correction for the $k$ identifiable directions.

BIC instead approximates the partition function
\begin{equation}
    Z_N=\int \exp[-N L_N(\theta)]\varphi(\theta)d\theta,
\end{equation}
where $F_N^{\mathrm{eq}}=-\ln Z_N$.
By the Laplace approximation,
\begin{equation}
    F_N^{\mathrm{eq}}
    \approx N L_N(\hat{\theta})+\frac{k}{2}\ln N+O(1).
    \label{eq:BIC}
\end{equation}
The first term is the empirical ground-state energy, while the second is the entropic cost from the posterior volume shrinking as $N^{-k/2}$.
Thus, AIC estimates predictive energy after correcting training bias, whereas BIC estimates the equilibrium Bayesian free energy of the posterior basin.

\textit{Singular models.---}
The thermodynamic intuition underpinning AIC and BIC breaks down in hierarchical Bayesian models and modern neural networks. 
In these systems, the mapping from parameters to probability distributions is not one-to-one, and the Fisher information matrix is often rank deficient. 
The potential landscape is not a simple quadratic basin but contains singularities: broad flat valleys, intersecting ridges, and manifolds of equivalent parameters. 
In such regions, the central limit theorem for the estimator does not hold in its regular form, and the posterior distribution is no longer Gaussian, even asymptotically.

To address this issue, Watanabe developed singular learning theory based on algebraic geometry \cite{Watanabe2009}. 
The fundamental insight is that Hironaka's resolution of singularities \cite{Hironaka1964,Hironaka1964b} allows us to transform the singular loss landscape into a normal-crossing form via a birational coordinate transformation. 
In the resolved coordinates, the posterior integral can be evaluated asymptotically. 
This process reveals that the effective number of parameters $k/2$ appearing in BIC should be replaced by the real log canonical threshold $\lambda$.
For regular models, $\lambda=k/2$; for singular models, $\lambda$ is generally smaller and reflects the algebraic geometry of the dominant singularity.

The singular counterpart of AIC 
is WAIC for predictive accuracy \cite{Watanabe2010}.
It replaces the regular equipartition correction by an empirical posterior fluctuation of the log likelihood.
Thus WAIC estimates the generalization loss without assuming a Gaussian posterior or identifiable parameters.

For Bayesian model selection, however, we need the model evidence, or equivalently the free energy.
This leads to WBIC \cite{Watanabe2013}.
In singular models, the equilibrium free energy has the asymptotic form
\begin{equation}
    F_{N}^{\mathrm{eq}}
    =
    N L_{N,\min}
    +\lambda\ln N
    -(m-1)\ln\ln N
    +O(1),
    \label{eq:singular_free_energy}
\end{equation}
where $m$ is the multiplicity of the dominant pole.
Compared with Eq.~(\ref{eq:BIC}), the regular penalty $(k/2)\ln N$ is replaced by the singular complexity $\lambda\ln N$, up to the logarithmic multiplicity correction.
Direct computation of $\lambda$ requires resolving the singularities of the model and is generally impractical for large learning machines.
WBIC avoids this algebraic computation by evaluating the internal energy at a special inverse temperature.

To define it, introduce the tempered partition function
\begin{equation}
    Z_N(\beta)
    =
    \int \exp[-\beta N L_N(\theta)]\varphi(\theta)d\theta,
\end{equation}
and the corresponding tempered expectation
\begin{equation}
    \mathbb{E}_{\beta}[A(\theta)]
    =
    \frac{1}{Z_N(\beta)}
    \int A(\theta)
    \exp[-\beta N L_N(\theta)]\varphi(\theta)d\theta.
\end{equation}
Near the dominant singularity, singular learning theory gives
\begin{equation}
    Z_N(\beta)
    \approx
    C(\beta N)^{-\lambda}
    \{\ln(\beta N)\}^{m-1}
    \exp[-\beta N L_{N,\min}],
    \label{eq:tempered_partition}
\end{equation}
where $C$ is independent of $N$ at leading order.
Differentiating $\ln Z_N(\beta)$ gives
\begin{equation}
    \mathbb{E}_{\beta}[N L_N(\theta)]
    \approx
    N L_{N,\min}
    +\frac{\lambda}{\beta}
    +O\!\left(\frac{1}{\beta\ln(\beta N)}\right).
    \label{eq:average_energy_scaling}
\end{equation}
At the inverse temperature $\beta^*=1/\ln N$, the thermal correction $\lambda/\beta$ becomes the free-energy penalty $\lambda\ln N$.
WBIC is therefore defined as
\begin{equation}
    \mathrm{WBIC}_{N}
    =
    \mathbb{E}_{\beta^*}[N L_N(\theta)].
\end{equation}
It estimates the equilibrium free energy in Eq.~(\ref{eq:singular_free_energy}) up to lower-order terms.
In particular, a Monte Carlo sampling at $\beta^*$ gives access to the free-energy penalty without explicitly calculating the RLCT.
This derivation, however, relies on the assumption that the sampling process has reached the equilibrium tempered posterior at $\beta^*$.

\textit{Non-equilibrium extension.---}
Now consider a realistic scenario where the search for $\hat{\theta}$, or the sampling of the posterior, is constrained by a finite resource $R$.
The resource may be the number of gradient steps, the number of MCMC updates, the annealing time, or a more general computational budget.
In this scenario, the parameter $\theta$ does not necessarily sample from the equilibrium posterior.
Instead, it evolves according to a stochastic dynamical process.
Let the sequence of parameters be $\theta_{0:T}=\{\theta_0,\theta_1,\ldots,\theta_T\}$.
The probability of realizing a specific trajectory is governed by the path probability
\begin{equation}
    P(\theta_T|R)
    =
    \int d\theta_0\cdots d\theta_{T-1}\,
    p(\theta_0)
    \prod_{t=1}^{T}K(\theta_t|\theta_{t-1}),
    \label{eq:path_probability}
\end{equation}
where the transition kernel $K$ depends on the learning rule and, when present, on the noise level used in the stochastic dynamics.
For concreteness, suppose that the learning algorithm is approximated by stochastic gradient Langevin dynamics \cite{Welling2011}.
A simple update is
\begin{equation}
    \theta_t
    =
    \theta_{t-1}
    -\eta\nabla L_N(\theta_{t-1})
    +\xi_t,
\end{equation}
where $\eta$ is the learning rate and $\xi_t\sim\mathcal{N}(0,2\eta\beta^{-1}N^{-1}I)$.
The corresponding transition kernel can be the Gaussian function $K_{\beta}(\theta_t|\theta_{t-1})$.
When $L_N(\theta)$ is real analytic, as in networks with analytic activation functions such as tanh or sigmoid, or in smoothed approximations to ReLU networks, then the Gaussian kernel is analytic in its arguments.
This observation motivates the use of an analytic finite-time effective potential.

There is a subtle but important point here.
It is too strong to claim that the convolution in Eq.~(\ref{eq:path_probability}) automatically preserves all the structures needed for singular learning theory for every algorithm and every noise model.
What we need is a local effective representation of the finite-time endpoint ensemble.
Under smooth noisy dynamics with nondegenerate diffusion, the endpoint density can often be written, at least locally and up to smooth prefactors, in the form
\begin{equation}
    q_R(\theta)
    =
    \frac{1}{Z_R(1)}
    \exp[-N L_R(\theta)]\varphi_R(\theta),
    \label{eq:effective_ensemble}
\end{equation}
where $L_R(\theta)$ is a finite-time effective loss and $\varphi_R(\theta)$ collects smooth contributions from initialization, noise, and the path measure.
Equivalently,
\begin{equation}
    L_R(\theta)
    =
    -\frac{1}{N}\ln q_R(\theta)
    +\frac{1}{N}\ln\varphi_R(\theta)
    +\mathrm{const.}
\end{equation}
This is the key modeling step of the non-equilibrium extension.
The construction is not a theorem for arbitrary learning dynamics, but it provides a controlled setting in which the endpoint ensemble can be analyzed using the same asymptotic framework as an equilibrium posterior.
More precisely, the effective loss $L_R$ exists locally whenever the endpoint density $q_R$ is positive and real analytic with respect to a positive real-analytic reference factor $\varphi_R$.  In the discrete-time SGLD example above, if $L_N$ is real analytic and the Gaussian noise is nondegenerate, each finite-step kernel is real analytic in both endpoints; for a finite horizon $T$, the endpoint density obtained by integrating a finite product of such kernels is locally analytic under the usual dominated-convergence conditions.  This assumption is the finite-time analogue of the analyticity assumption in singular learning theory.  For continuous-time diffusion limits, standard results on elliptic diffusions give smooth transition densities under nondegeneracy conditions \cite{KusuokaStroock1985}; analyticity is an additional regularity assumption on the effective endpoint ensemble.
For the WBIC construction, we then introduce a formal tempered family generated from this fixed effective loss,
\begin{equation}
    Z_R(\beta)=
    \int \exp[-\beta N L_R(\theta)]\varphi_R(\theta)d\theta.
\end{equation}
This convention avoids differentiating through an additional, algorithm-dependent temperature dependence of the endpoint density itself.

If the effective loss $L_R(\theta)$ is real analytic in the neighborhood that dominates the endpoint distribution, Hironaka's resolution of singularities can be applied to this finite-time landscape.
The resulting singularity is characterized by a resource-dependent RLCT, denoted by $\lambda_R$, and a multiplicity $m_R$.
The corresponding finite-time free energy ($F_R=-\ln Z_R(1)$) is
\begin{equation}
    F_R
    \approx
    N L_{R,\min}
    +\lambda_R\ln N
    -(m_R-1)\ln\ln N
    +O(1).
    \label{eq:noneq_free_energy}
\end{equation}
Here $L_{R,\min}$ is the effective energy level reached or resolved by the finite-time process, not necessarily the global minimum of the original empirical loss.
The parameter $\lambda_R$ measures the singular complexity of the region in the finite-time dynamics.

We can now repeat the WBIC argument in this finite-time ensemble.
The tempered endpoint partition function has the same leading form as Eq.~(\ref{eq:tempered_partition}), with $L_N$ and $\lambda$ replaced by $L_R$ and $\lambda_R$:
\begin{equation}
    Z_R(\beta)
    \approx
    C_R(\beta N)^{-\lambda_R}
    \{\ln(\beta N)\}^{m_R-1}
    \exp[-\beta N L_{R,\min}].
\end{equation}
Therefore,
\begin{equation}
    \mathbb{E}_{R,\beta}[N L_R(\theta)]
    \approx
    N L_{R,\min}+\frac{\lambda_R}{\beta}
    +O\!\left(\frac{1}{\beta\ln(\beta N)}\right).
    \label{eq:noneq_average_energy}
\end{equation}
At the WBIC temperature $\beta^*=1/\ln N$, this becomes
\begin{equation}
    \mathrm{WBIC}_{R}
    \equiv
    \mathbb{E}_{R,\beta^*}[N L_R(\theta)]
    \approx
    N L_{R,\min}+\lambda_R\ln N.
    \label{eq:noneq_wbic}
\end{equation}
This expression is the finite-time analogue of WBIC.
It captures two effects simultaneously: the achieved energy level of the finite-time learner and the singular complexity of the finite-time region explored by that learner.

The complexity $\lambda_R$ should not be interpreted as a static architectural constant.
It depends on the resource $R$, the algorithm, the temperature schedule, and the region of parameter space reached by the dynamics.
In early stages of learning, the endpoint distribution may only resolve a broad, effectively simple basin.
At later stages, the dynamics may access finer singular structures of the model.
However, no monotonicity of $\lambda_R$ should be assumed in general: a learning trajectory may move between basins, become trapped in metastable regions, or concentrate on lower-dimensional structures.
If the dynamics mix to the equilibrium posterior in the limit $R\to\infty$, then $L_R$ and $\lambda_R$ approach their equilibrium counterparts.
This viewpoint is closely related to the local learning coefficient (LLC), recently introduced as a singularity-aware complexity measure during training \cite{Lau2025LLC,Wang2025RLLC}.  
The LLC is based on Watanabe's volume characterization of the RLCT near the ground state \cite{Watanabe2009} and provides a practical SGLD-based route to estimating a local RLCT around a local minimum.  
This suggests a corresponding way to estimate $\lambda_R$: restrict the LLC-type volume or tempered-sampling estimate to the endpoint ensemble, or to neighborhoods sampled from the finite-resource dynamics specified by $(K,T)$.  The difference is that the usual LLC is estimated locally around selected parameter values, often along an optimization path, whereas $\lambda_R$ is an ensemble-level effective coefficient depending on the algorithm, time horizon, and temperature schedule.  
Both quantities share the same geometric origin and are naturally connected with Watanabe's singular learning process.

We can now clarify the relationship with epiplexity \cite{Finzi2026}.
The finite-time free energy in Eq.~(\ref{eq:noneq_free_energy}) is not epiplexity itself; rather, it gives the thermodynamic contribution to the time-bounded MDL objective for a fixed model class and learning protocol.
In the terminology of Ref.~\cite{Finzi2026}, $N L_{R,\min}$ is entropy-like, whereas $\lambda_R\ln N$ is a structural complexity term within that class.
Epiplexity is obtained after the time-bounded MDL minimization selects the structural part over possible predictors.
Thus a natural finite-resource MDL criterion is
\begin{equation}
    \mathcal{C}_R(D_N)
    =
    \min_{M:\,\mathrm{Cost}(M)\le R}
    \left\{
    \mathrm{WBIC}_{R}(M)+|M|
    \right\},
    \label{eq:finite_resource_mdl}
\end{equation}
where $|M|$ denotes the explicit code length of the model class, architecture, or learning protocol.
Here, $\mathrm{WBIC}_{R}(M)$ estimates the stochastic thermodynamic cost for a fixed candidate, whereas the minimization over $M$ implements the MDL step from which epiplexity extracts the structural component.

The main value of our result is calculability.
WBIC turns the model-conditioned stochastic complexity into a thermal expectation.
We extend this idea to finite-resource learning by replacing the equilibrium RLCT with the effective quantity $\lambda_R$.
Watanabe's theory thus bridges the time-bounded code length of epiplexity and a quantity that can be estimated by sampling or learning dynamics.
The ordinary criteria also compare models after assuming that training or sampling has reached the relevant equilibrium object.
The finite-time criterion instead compares what a real learning procedure can generate under a specified resource.
It can distinguish models with similar equilibrium evidence but different transient accessibility, or algorithms applied to the same architecture but reaching different effective singularities.
The optimal model under finite computation is therefore the model--algorithm pair with the shortest attainable description length under the given resource constraint.

The author acknowledges financial support from the Cross-ministerial Strategic Innovation Promotion Program (SIP) of the Cabinet Office (No.~23836436).
\bibliographystyle{apsrev4-2}
\bibliography{references}

\end{document}